\documentclass[12pt]{article}
\pdfoutput=1
\usepackage[margin=1in]{geometry}
\usepackage{amsmath}
\usepackage{amssymb}
\usepackage{hyperref}
\usepackage{graphicx}
\usepackage{bm}
\usepackage[sort&compress,numbers]{natbib}
\def\v{\bm{v}}
\def\z{\bm{z}}
\def\E{\bm{E}}

\def\P{\bm{P}}
\def\eps{\bm{\epsilon}}
\def\I{{\bf I}}
\title{Oblate Electron Holes are not attributable to Anisotropic Shielding}
\author{I H Hutchinson}
\date{Plasma Science and Fusion Center,\\ Massachusetts Institute of
  Technology,\\ Cambridge, MA, USA.}

\begin{document}
\maketitle

\begin{abstract}
  Shielding mechanisms' influence on the ratio of perpendicular to
  parallel scale lengths of multidimensional plasma electron hole
  equilibria are analyzed theoretically and computationally. It is
  shown that the ``gyrokinetic'' model, invoking perpendicular
  polarization, is based on a misunderstanding and cannot explain the
  observational trend that greater transverse extent accompanies lower
  magnetic field. Instead, the potential in the wings of the hole,
  outside the region of trapped-electron depletion, has isotropic
  shielding giving $\phi\propto {\rm e}^{-r/L}/r$, with the shielding
  length $L$ equal to the Debye length for holes much slower than the
  electron thermal speed. Particle in cell simulations confirm the
  analysis.
\end{abstract}

\section{Introduction}

Plasma electron holes\cite{Hutchinson2017} are widely observed by
satellites in space.  They are a major subset of the various observed
Electrostatic Solitary Wave (ESW) structures that are important
elements of the plasma turbulence in various regions. They move along
the magnetic field ($B$) at speeds a fraction of the electron thermal
speed, thus suppressing the (slower) ion response (which will be
ignored here), and are sustained by a deficit of electrons on trapped
orbits: a kind of Bernstein-Greene-Kruskal (BGK) mode. Although
one-dimensional BGK analysis\cite{Bernstein1957} seems to provide an
understanding of the parallel (to $B$) structure, electron holes are
frequently observed to be two- or three-dimensional: of limited
transverse extent. It has been shown computationally
\cite{Muschietti2000,Wu2010} and explained with quantitative analysis
\cite{Hutchinson2018a,Hutchinson2019,Hutchinson2019a} that so-called
transverse instabilities break up one-dimensional holes into smaller
transverse length scales. And, increasingly, multi-satelite (e.g.\
Cluster\cite{Graham2016} and MMS\cite{Steinvall2019,Lotekar2020})
observations are documenting the transverse potential structure. It
therefore seems essential to develop a multi-dimensional understanding
of their equilibria and stability, in order to understand the
observations and simulations. The purpose of this paper is to rule out
from consideration a much cited but erroneous hypothesis proposed to
explain the multi-dimensional shape of electron holes; and hence to
indicate more narrowly what mechanisms could be important.

In their pioneering observations, using the Polar satellite, Franz et
al\cite{Franz2000} represented the shape of electron holes by the
ratio of scale lengths $L_\perp/L_\parallel$, perpendicular and
parallel to the magnetic field. They found that between $1.8$ and $9$
earth radii, weaker magnetic field regions of the magnetosphere had
holes that were statistically more oblate: more elongated in the
directions perpendicular to the magnetic field. They fitted their
results as $L_\perp/L_\parallel \simeq \sqrt{1+\omega_p^2/\Omega^2}$,
where $\omega_p$ is the plasma frequency, and $\Omega$ the (electron)
cyclotron frequency. They also offered a speculation that the reason
for this scaling is to be found in a ``modified'' Poisson equation of
the form
\begin{equation}
  \label{eq:PoissonMod}
  \left[\nabla_\parallel^2+(1+{\omega_p^2\over\Omega^2})\nabla_\perp^2\right]
  \phi = -\rho_s/\epsilon_0,
\end{equation}
which in the context of the ``gyrophase averaged Vlasov equation'',
arises ``to include the polarization density of the particles''. They
referenced early papers\cite{Zakharov1974,Lee1983} relating to what is now called gyrokinetics, a
reduced model much used for low frequency turbulence studies. This
explanation has been taken up by several subsequent authors in electron
hole research\cite{Jovanovic2002a,Berthomier2003,Wu2010,Vasko2017,Vasko2018,Holmes2018,Tong2018,Fu2020}, often being approvingly called the gyrokinetic scaling
relation.

If this model really applied to electron holes, its effects would
be a vital part of multi-dimensional equilibria. However, it does not,
because it misunderstands gyrokinetics. The phenomenon of
transverse polarization \emph{cannot provide the explanation} of the
observed hole aspect ratio scaling because its contribution is always
small, as is now explained.

\section{Transverse plasma polarization}

\subsection{Elementary derivation}
Transverse polarization is discussed in text books.  Addressing just
the motion of electrons (charge $q$, mass $m$, density $n_e$), the
particles drift in response to perpendicular electric fields via the
$\bm v_{E\times B}=\bm E\times \bm B/B^2$ drift and, in a time
dependent $\bm v_{E\times B}$, polarization drift:
$\v_p={m\over qB^2} \dot\E_\perp=-{m\over qB^2}\nabla_\perp\dot\phi$
arises from inertial force.  Integrating $\bm v_p$ with respect to
time, the displacement gives an electric polarization density
\begin{equation}
  \label{eq:polariz}
  \P=qn_e\int\v_p dt=-{n_em\over B^2}\nabla_\perp \phi =-\epsilon_0{\omega_p^2\over \Omega^2}\nabla_\perp \phi  
\end{equation}
where $\Omega=|q|B/m$ and $\omega_p^2=n_eq^2/\epsilon_0m$.

If, then, there exists a specified charge density $\rho_s$ that
somehow excludes the electron polarization response, so in terms
of elementary dielectric theory it is a ``free-charge-density'', yet
resides in an electron plasma in which \emph{the only electron
  response is the polarization drift}, the potential will be governed
by Poisson's equation containing, in addition to $\rho_s$, a
polarization charge density
$\rho_\varepsilon=-\nabla.P=
\epsilon_0{\omega_p^2\over\Omega^2}\nabla_\perp^2 \phi$. Hence,
Poisson's equation
$-\nabla^2\phi =(\rho_s+\rho_\varepsilon)/\epsilon_0$ can be written
in the modified form of equation (\ref{eq:PoissonMod}):
$
\left[\nabla_\parallel^2+(1+{\omega_p^2\over\Omega^2})\nabla_\perp^2\right]
\phi = -\rho_s/\epsilon_0$.  This is equivalent to saying that the
polarization drift gives rise to an anisotropic (relative) dielectric
tensor $\eps=\I+(1+\omega_p^2/\Omega^2)(\I-\hat{\z}\hat{\z})$, where
$\hat\z$ is the magnetic field direction.

Some electron hole
papers suppose
that the anisotropy should be scaled away by defining new transverse
spatial coordinates $(x',y')=(1+\omega_p^2/\Omega^2)^{-1/2}(x,y)$ leading
to isotropic solutions in the new coordinates, depending only on
$r'=\sqrt{x'^2+y'^2+z^2}$, which on this assumption would give Franz's
$L_\perp/L_\parallel$ scaling. But this is a mistake because the
treatment has implicit approximations not
satisfied for electron holes, and ignores the general properties of
untrapped collisionless orbits in a potential well.

\subsection{The dielectric response in gyrokinetics}

The gyrokinetic approximation is based on an ordering that takes the
perturbation to be slowly varying both intrinsically as
$|{\partial\over \partial t}| \simeq \omega \ll \Omega$ and
convectively along the magnetic field as $k_\parallel v_t\ll
\Omega$. Gyrokinetics' strength (in modern formulations) is that (unlike the
derivation just given and the references cited by Franz) it does not
necessarily assume the gyro-radius $r_L$ to be small compared with the
transverse perturbation scale. For $k_\perp r_L \gtrsim 1$ one must
average over the gyroradius, as if the gyrating particle were a ring
of charge on its approximately circular transverse orbit. Also,
since gyrokinetics is generally expressed as the motion of the
gyrocenter, its Poisson equation (for an electrostatic problem)
requires a transformation from gyrocenter-density, back to
particle-density, which is what determines the local charge density
and potential. This is where the polarization drift enters.
{Actually gyrokinetic analyses often replace Poisson's
  equation with quasineutrality, and address the ion polarization more
  than the electron polarization, neither of which is appropriate for
  electron hole analysis. The electron polarization is the question
  here, but is to be dealt with using the same analysis as in standard
  gyrokinetics for the density of the ions.}

A helpful discussion of the resulting transverse dielectric constant
and its relationship to what is known about the full dielectric
response without gyrokinetic approximations is found in reference \cite{Krommes1986}. From that paper the electron contribution important
for holes can be drawn by extension of their ion analysis. The
effective electron gyrokinetic transverse (relative) dielectric
constant expressed in Fourier space is
\begin{eqnarray}
  \label{eq:DC1}
  \epsilon_\perp  &=&  1+{1\over k^2
    \lambda_{De}^2}(1-\Gamma_0(b))\\
  \label{eq:DC2}
  &\simeq& 1+\left(k_\perp\over k \right)^2\left(\omega_{pe}\over \Omega\right)^2
  \qquad (b = k_\perp^2r_L^2\ll 1)\\
  \label{eq:DC3}
  &\simeq& 1 + \left(\omega_{pe}\over \Omega\right)^2
  \qquad\qquad (k_\parallel/k_\perp\ll 1),
\end{eqnarray}
where $\lambda_{De}$ is the electron Debye length, and
$\Gamma_0(b) = {\rm e}^{-b}I_0(b)$, with $b\equiv k_\perp^2T/\Omega^2m$
the thermal value of $k_\perp^2r_L^2$.  A small-$b$ expansion has
$\Gamma_0(b)\simeq 1-b$, which is the basis for the first
approximation in this sequence [eq.\ (\ref{eq:DC2}) noting that
$\omega_{pe}\lambda_{De}=\sqrt{T/m}$]. However using it for
$b\gtrsim 1$ gives completely the wrong dependence since as
$b\to \infty$, $\Gamma_0(b)\to 1/(2\pi b)^{1/2}$ remaining positive.
A somewhat better (Pad\'e) approximation\cite{Hammett1991}
$\Gamma_0(b)\simeq 1/(1+b)$, avoids so egregious an error, giving
$1-\Gamma_0\to 1$ as $b\to \infty$, the unmagnetized result.  The
electron hole elongation model explanation requires eq.\
(\ref{eq:DC3}) to apply into the regime where
$\omega_{pe}/ \Omega =r_L/\lambda_{De} \gg 1$. Since electron holes
have $k\sim 1/\lambda_{De}$, it never does. Instead, at weak magnetic
field the large-$b$ limit applies: $\epsilon_\perp \to 1$.

Summarizing: the so-called gyrokinetic model to explain electron hole
transverse extent is mistaken because in the regime
$\omega_{pe}\gtrsim \Omega$, where the polarization term is relevant,
(1) gyrokinetics does not apply to electron holes because their
parallel length is too short ($k_\parallel v_t\not\ll \Omega$);
(2) Franz's expression (\ref{eq:DC3}) uses an approximate form for the
gyrokinetic dielectric response in a regime where it is invalid; and
(3) polarization drift is not the only electron response, as will now
be discussed.

\section{Density in a general attractive potential}

An independent demonstration of the point can be derived from far more
general considerations, and these also bear directly on the existence
of multidimensional hole equilibria.  A collisionless
multi-dimensional electron hole (attractive electrostatic potential)
time-invariant in some frame of reference moving along the uniform
magnetic field, has electron populations that are either trapped or
untrapped. The effect of an electrostatic structure like this on the
density of the attracted species is a general matter long studied for
electric probes and other plasma perturbing
bodies\cite{Alpert1965,Laframboise1966a,Laframboise1993}; the present
case differs only in that there is no particle absorption by a body
--- a simplification. Any point in phase-space is on either a trapped
or untrapped orbit. 

Henceforth we use normalized units $\lambda_{De}$ for length,
$1/\omega_{pe}$ for time, the background temperature $T_{e\infty}$ for
energy, and express densities normalized to the distant background
electron density $n_{e\infty}$; the electron mass is then unity and
charge $-1$. In steady state, the total energy $W=v^2/2-\phi$
(normalized units) of a particle is conserved, and in collisionless
situations, the distribution function is conserved along
orbits. Therefore on any untrapped orbit at potential $\phi$ the (3-D)
distribution function is $f(v)=f_\infty(v_\infty)$, the distant
(assumed uniform) distribution $f_\infty$ at the velocity
$v_\infty=\sqrt{2W}$ ($\phi_\infty=0$). All untrapped orbits will be
populated at this level, and if the trapped phase-space is negligible,
a simple Boltzmann factor density dependence is the outcome. The
\emph{trapped} orbits' distribution, by contrast, is not determined in
this way by boundary conditions, and requires consideration of initial
conditions or weak collisions\cite{Lampe2001,Hutchinson2007a}.

In cylindrical or spherical symmetry situations, determining which
orbits are trapped is considerably complicated by the conservation of
corresponding angular momentum or $z$-momentum (for
$z$-ignorable). However, in a multidimensional electron hole, in which
cylindrical angle $\theta$ is the only possibly ignorable coordinate,
the only exactly conserved quantity in addition to energy is the
canonical angular momentum $p_\theta$ which mostly restricts the
radial ($x^2+y^2$) excursion, and does not prevent orbits from
escaping the potential well along the magnetic field, in the
$z$-direction. Therefore orbits of positive total energy ($W$) are trapped
only when the magnetic moment ($\mu=mv_\perp^2/2B$) is adiabatically
conserved, and the resulting conservation of parallel energy
$W_\parallel={1\over2} v_\parallel^2-\phi$ causes parallel trapping
if $W_\parallel$ is
negative. When $\mu$ conservation is violated because the gyroradius
is comparable to the hole's size, the parallel trapping
begins to break down and eventually the only trapped particles are
those with negative total energy $W$.

If the \emph{only trapped} orbits are those with negative \emph{total}
energy $W=v^2/2-\phi<0$, then the untrapped density is
$n_u=\int \int_{v_m}^\infty f_\infty(v_\infty) v^2dv\, d\Omega_s$,
where $v_m=\sqrt{2\phi}$ and $d\Omega_s$ is the solid angle element.
For an unshifted 3-D Maxwellian distant distribution,
$f_\infty(v)=(2\pi)^{-3/2}\exp(-v_\infty^2/2)$, the untrapped density
expression can be integrated by parts using $v_\infty^2=v^2-v_m^2$:
\begin{equation}
\begin{split}
  n_{3u}&={4\pi\over(2\pi)^{3/2}}\int_{v_m}^\infty
  \exp([v_m^2-v^2]/2)v^2dv
  ={2\over (2\pi)^{1/2}}\left\{v_m+ \int_{v_m}^\infty \exp(-v_\infty^2/2)dv\right\}.
\end{split}
\end{equation}
This should be compared with the 1-D density of particles that are not
\emph{parallel} trapped when $v_\perp$ and
${1\over2}v_\parallel^2-\phi$ are conserved, those with
$W_\parallel>0$, which is:
\begin{figure}[htp]
  \centering
  \includegraphics[width=0.5\hsize]{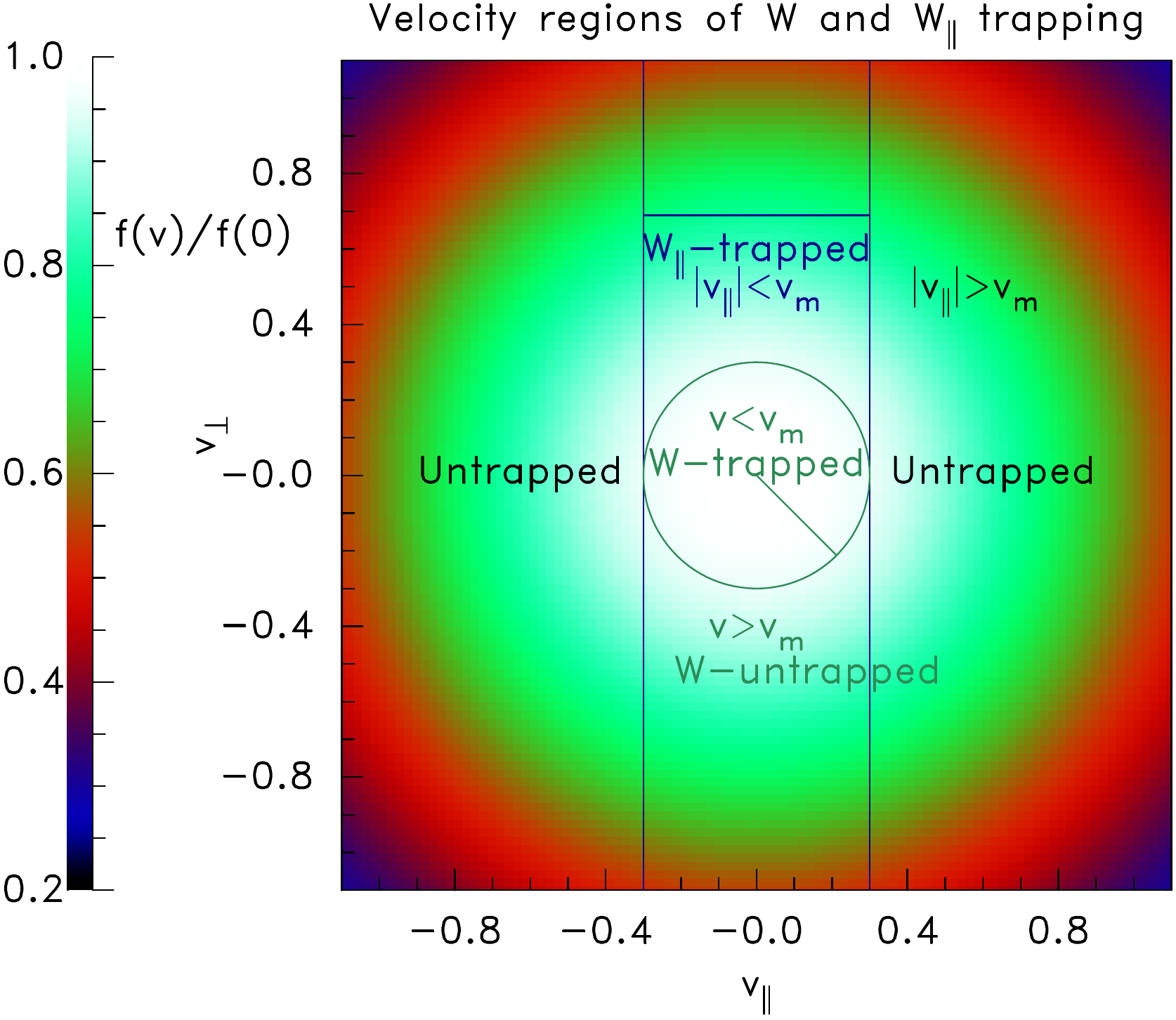}
  \caption{Schematic of the total-energy-trapped and parallel-trapped
    regions in velocity space.}
  \label{fig:trapreg}
\end{figure}
\begin{equation}
  \label{eq:1-Dpassing}
  n_{1u}= {2}\int_{v_m}^\infty f_\parallel dv_\parallel=
  {2\over (2\pi)^{1/2}}\int_{v_m}^\infty\exp(-v_{\parallel\infty}^2/2)dv_\parallel.
\end{equation}
The velocity domains of these two types of untrapped density are
illustrated in Fig.\ \ref{fig:trapreg}.  There is far more phase space
volume subject to parallel trapping (when $W_\parallel$ is conserved)
than to total energy trapping.  The difference in the densities,
$(n_{3u}-n_{1u})=2v_m/\sqrt{2\pi}$, is exactly equal to the
contribution from the ($W_\parallel$-)trapped region to density in a
one-dimensional hole when the trapped distribution is flat
$f_{\parallel t}=f_{\parallel\infty}(0)$. We denote the total density
(trapped plus untrapped) of this flat-trapped distribution as
$n_{1f}$, and have $n_{3u}=n_{1f}$. It can easily be shown that
$n_{3u}$ and $n_{1f}\simeq 1+\phi$ to lowest order in $\phi$. (For
\emph{shifted} Maxwellian distributions the coefficient of $\phi$
becomes gradually smaller, the shielding weaker\cite[Fig
10]{Hutchinson2017}, but the generalization is straightforward, and
the shielding remains isotropic.) Therefore, even if the entire $W<0$
trapped velocity-region were completely depleted of electrons (making
$n_{3u}$ the total density) that would be insufficient to make the
electron density \emph{decrease with $\phi$} or give a positive charge
density for small $\phi$. This is the qualitative
reason\cite{Krasovsky2004,Hutchinson2017} why multidimensional
electron holes cannot exist without a magnetic field.  Any density
contribution (or deficit) from the $W<0$ region of phase space is of
the form
$n_{3t}=4\pi v_m^3\langle f_t\rangle/3=4\pi (2\phi)^{3/2}\langle
f_t\rangle/3$, which enters as a higher power $\phi^{3/2}$ than in
$n_{3u}-1(\propto \phi)$, justifying 3D Debye shielding ($n-1=\phi$)
in the small $\phi$ limit, regardless of trapped density.\footnote{The
  velocity integral of a two-dimensional Maxwellian over
  $(v_x^2+v_y^2)>2\phi$ gives untrapped density exactly $n_{2u}=1$,
  and the flat-trapped (integral over $(v_x^2+v_y^2)<2\phi$) density
  is $n_{2t}=\phi$, again giving a Boltzmann response
  $n_{2f}= (1+\phi)$.}

In summary, an assumption that orbits are trapped if and only if the
total energy is negative, $v^2/2-\phi<0$, gives rise, at lowest
order, to a Debye shielding density $n\simeq 1+\phi$, with an
isotropic Poisson equation $\nabla^2\phi=\phi$, and no additional
$(\omega_p/\Omega)^2$ term, regardless of trapped distribution. This
situation is the anticipated result of small magnetic field, for which
the gyro-radius $r_L\sim v_\perp/\Omega$ is much bigger than the
potential structure extent $L$.

On the other hand strong magnetic field, $r_L\ll L$ with
$L\sim \lambda_{De}$ makes the term
$(\omega_p/\Omega)^2=(r_L/\lambda_D)^2\sim (r_L/L)^2$ negligibly
small, again yielding an isotropic effective Poisson equation.  In
this limit parallel trapping predominates, and Debye shielding is
recovered only by taking the shielding distribution to include a flat
contribution $f_{\parallel t}=f_\infty(0)$ in the trapped velocity
region: in other words the electron deficit that sustains the hole is
the difference $f_{\parallel}-f_{\parallel\infty}(0)$ in the trapped
phase-space.

In the intermediate regime, $(\omega_p/\Omega)^2\sim 1$, whether an
orbit is or is not trapped cannot be decided simply, because magnetic
moment approximate conservation depends upon subtleties of orbit
velocity. But the trapped region in phase-space is certainly
intermediate between the two extremes of total-energy and
parallel-energy trapping. Therefore it can hardly be expected that the
untrapped density will be suppressed significantly below
$1+e\phi/T$, as would be implied by use of eq.\
(\ref{eq:PoissonMod}). These conclusions are consistent with our
discussion of gyrokinetics.

\section{Particle in Cell Confirmation}

\begin{figure}
  \centering
  \includegraphics[width=0.5\hsize]{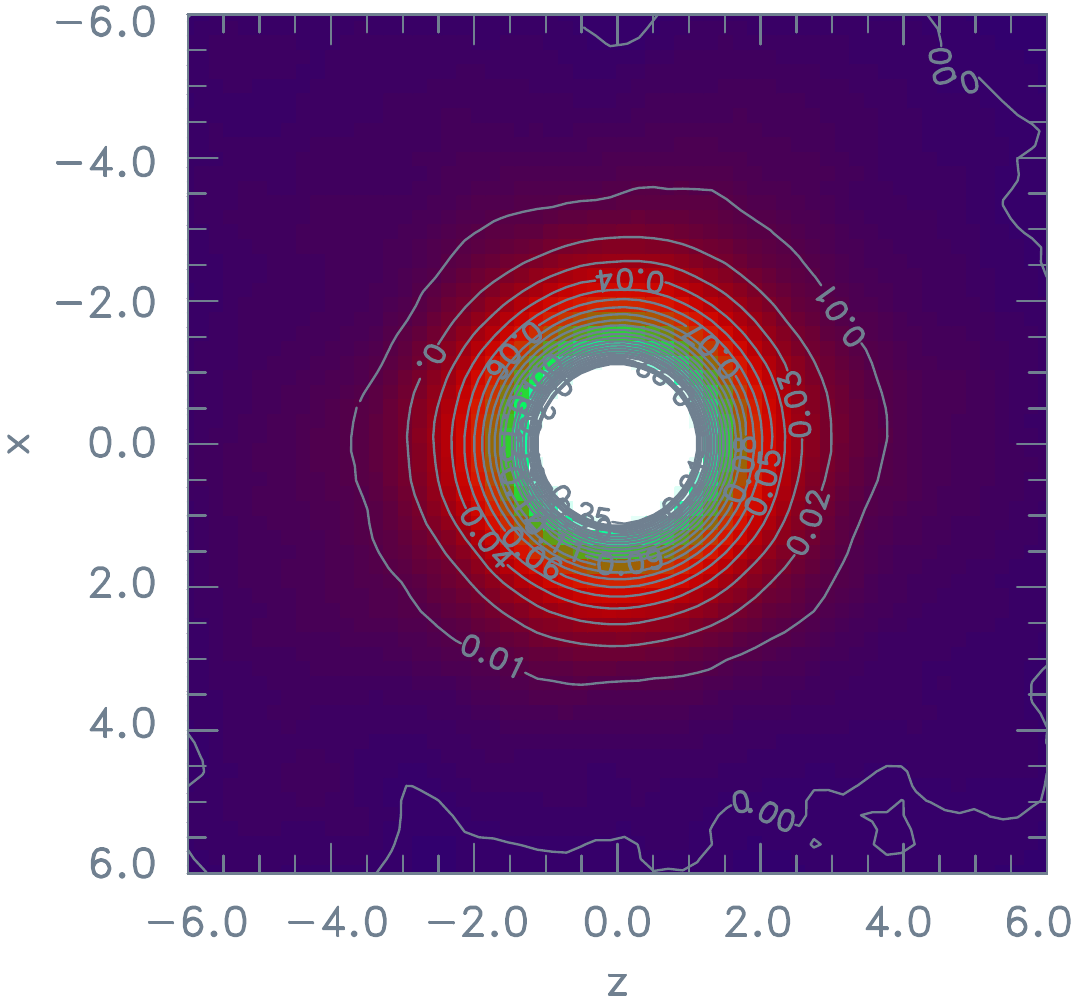}
  \caption{Potential contours around a non-absorbing equipotential surface (of
    unit radius and potential) show no significant anisotropy in a
    PIC simulation with $\omega_p/\Omega=5$.}
  \label{fig:circ2}
\end{figure}

A simple way to confirm the analytic conclusions arrived at here is to
perform full orbit, 6-dimensional phase-space, particle in cell (PIC)
simulations. The COPTIC code\cite{Hutchinson2011a} is used which is
capable of imposing a potential boundary condition embedded within the
plasma.  A cube of plasma with sides extending to $\pm 6$ (Debye
lengths) is used with a uniform magnetic field in the
$z$-direction. Potential is represented on a $60^3$ mesh with domain
boundary conditions $\partial \phi/\partial |x|=-\phi$ (and similarly
for $y$ and $z$) although the exact form is not observed to be
important.  The ($\sim 4$ million) particles move periodically at the
transverse boundaries $x$, $y$, but the $z$ boundaries are open with
particles reinjected representing an external Maxwellian. On a sphere
of radius 1, the potential is set to 1 (times $T_e/e$), representing
essentially a non-plasma charge distribution whose shielding by the
plasma is our interest. Particles move through this sphere without
being absorbed. The simulation is initialized with uniform electron
density and an equal and opposite fixed background charge density
representing ions; there is therefore a full complement of trapped and
untrapped particles, and the initial neutrality gives an initial
potential $\propto 1/r$. The code is advanced forward in time with
timesteps short enough to resolve the gyro-motion until a steady state
is reached. In this final state the potential is shielded with
characteristic length equal to $\sim 1$ (times $\lambda_D$). An
example contour plot of potential in the plane $y=0$ is shown in Fig.\
\ref{fig:circ2}.
It is almost exactly isotropic to within the noise level of the
simulation ($\sim 0.01$), showing no sign of transverse elongation
that would be present if anisotropic shielding such as eq.\
(\ref{eq:DC3}) applied.  The extensive explored range of values
$0.2<\Omega/\omega_p<2$ shows no tendency toward transverse elongation. Even
oblate elliptical inner boundary shapes which enforce elongation at
$r\sim1$ have potential contours at larger radii that rapidly become
spherical.

\section{Asymptotic potential variation of an electron hole}
\label{helmholtz}

We are now in a position to draw important conclusions about the form
of the potential in the wings of an electron hole. The wings are what
we call the spatial regions far enough from the hole center that the
trapped particle deficit (meaning difference from ``flat-trapped''
distribution) is negligible. The crucial point is then that in the
wings the density can be approximated as $n=1+\phi/\lambda_s^2$ (with
no coordinate anisotropy), where $\lambda_s$ is the effective
shielding length (for stationary Maxwellian external distribution the
Debye length, so $\lambda_s=1$, but longer for fast-moving
Maxwellians). The nonlinearity is weak because the potential is small
so $\lambda_s$ can be taken constant. Then the potential obeys the
Modified Helmholtz Equation
\begin{equation}
  \label{eq:modhelm}
  \nabla^2\phi-\phi/\lambda_s^2=0.
\end{equation}
In other places
where there is a trapped particle deficit, the resulting charge
density perturbation enters as a source on the right-hand side.

One can represent the solution of the Modified Helmholtz equation in
an infinite domain (isolated hole) as a sum of harmonics in the polar
angle $\theta=\tan^{-1}(r/z)$. For a 2-D cartesian case where $x$ is
an ignorable coordinate, $r$ corresponds to $y$, and $R^2=y^2+z^2$.
The harmonics then are proportional to modified Bessel functions
${\rm e}^{i\ell\theta} K_\ell(R/\lambda_s)$, whose asymptotic behavior is
$K_\ell(\eta) \to \exp(-\eta)\sqrt{\pi/2\eta}$ for large argument
$\eta$. For the cylindrical alternative (axisymmetry about $z$) the
harmonics are proportional to modified Spherical Bessel functions of
argument $\eta=R/\lambda_s=(z^2+r^2)/\lambda_s$ (where $r^2=x^2+y^2$), which are
asymptotically proportional to
$ \sqrt{\pi/2\eta}K_{\ell+1/2}(\eta)=\exp(-\eta)\pi/2\eta$, and for
$\ell=0$ simply the familiar Yukawa potential. The mix of angular
harmonics is determined by the angular dependence of the sources near
$R=0$, but the crucial point is that all of them decay radially predominantly
as $\exp(-\eta)$; so ${\partial \phi\over\partial R}\simeq -\phi/\lambda_s$. The
\emph{angular} component of the gradient is $\phi\ell/R$ which
for low $\ell$ and large $R$ becomes much smaller than
${\partial \phi\over\partial R}$.  Consequently, far from the charge
sources in the hole, when $R\gg \lambda_s$ ($\eta\gg 1$) the potential decays
mostly \emph{radially} and $\propto \exp(-R/\lambda_s)$.

For a one-dimensional hole or a multidimensional hole near $r=0$, the
asymptotic parallel variation is $\phi\propto \exp(-z/\lambda_s)$ at
large $|z|$; but also for any multidimensional hole shape, the
asymptotic transverse variation is $\phi\propto \exp(-r/\lambda_s)$
near $z=0$. These are physically inescapable asymptotes where trapped
particles are negligible.\footnote{Gaussian potential
  shapes are frequently used in hole modelling, for mathematical
  convenience; and the choice is often justified as consistent with
  observations. However the observations so far reported have
  insufficient precision in the wings to distinguish between
  exponential and Gaussian asymptotic decay.}

\begin{figure}[htp]
  \includegraphics[width=0.45\hsize]{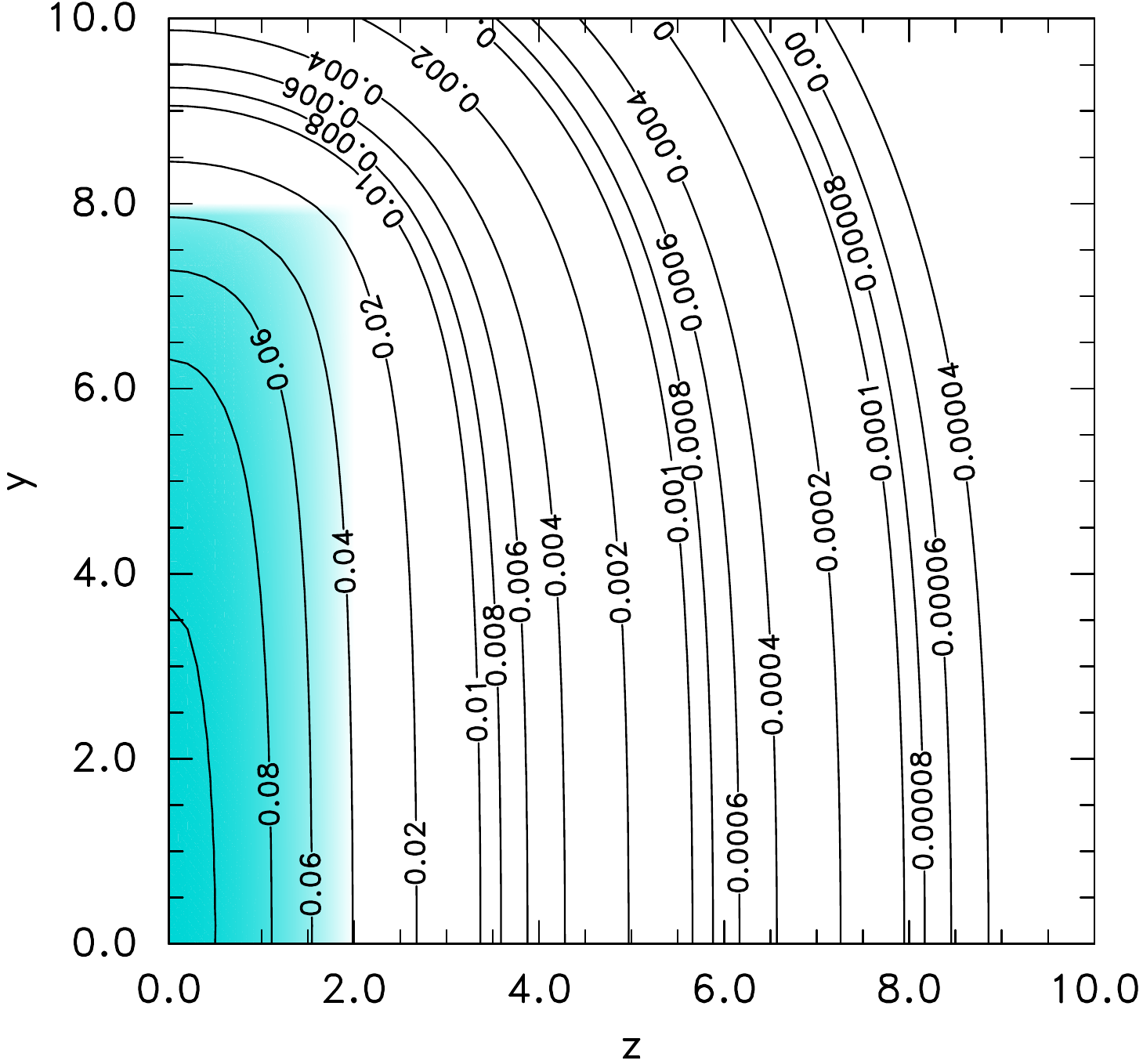}\hskip-1em(a)
  \includegraphics[width=0.45\hsize]{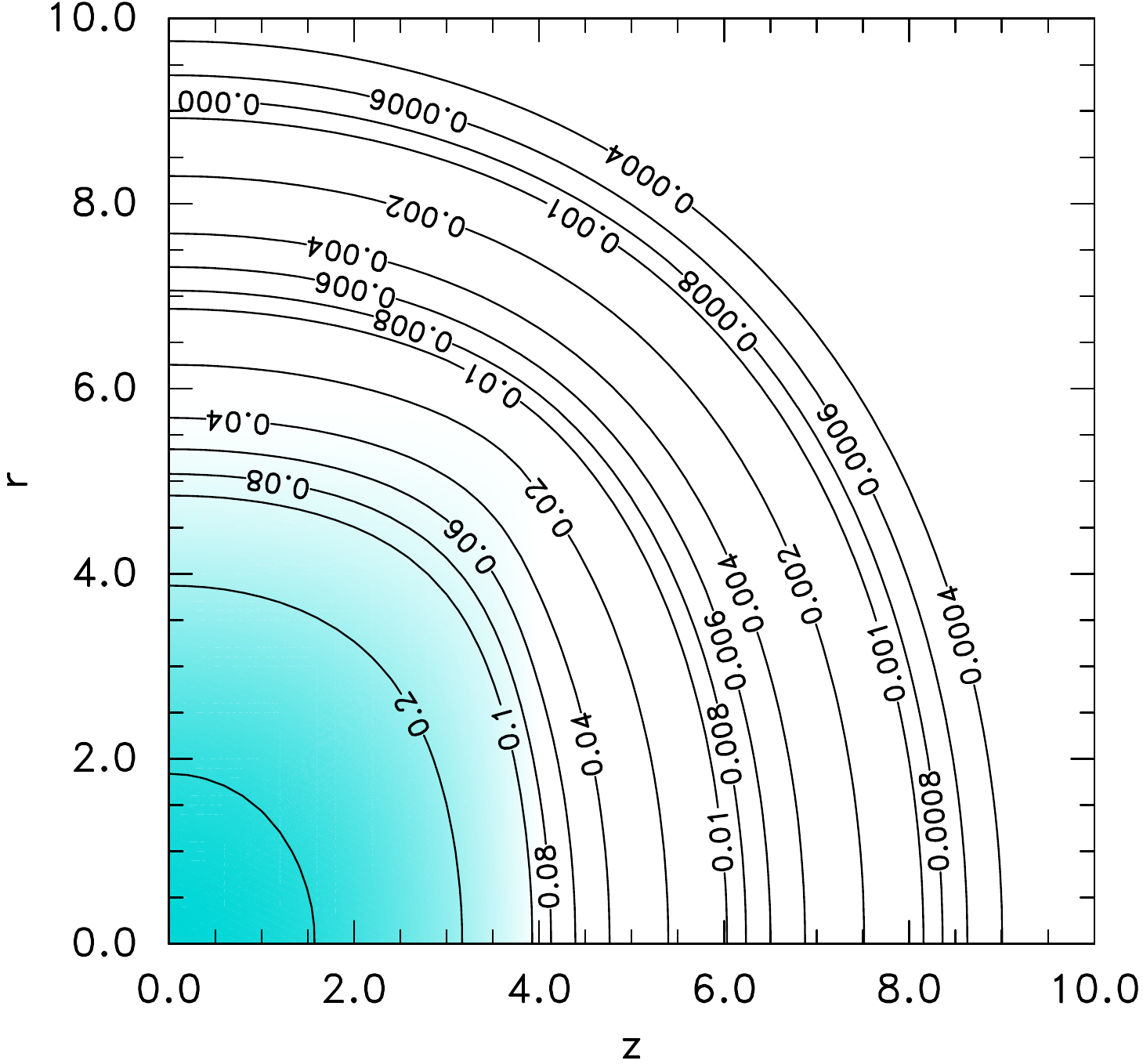}\hskip-1em(b)

  \caption{Examples of hole-like solutions of the Modified Helmholtz
    equation.}
  \label{fig:helmsol}
\end{figure}
Figure \ref{fig:helmsol} shows two examples of numerical solutions of
the Modified Helmholtz equation on 2-D cartesian (left) (a), and
axisymmetric (right) (b) domains with $\lambda_s=1$. A specified additional
charge source, representing trapped electron deficit (schematically,
not self-consistently by actual trapping), is indicated by the blue
color intensity on $\phi$ contour plots.  Notice how
the logarithmic $\phi$ contours are spaced (outside the shaded region)
by the same distance at $z\sim 0$ as they are at $r,y\sim 0$, and that
as we move further from the charge region the contours become
increasingly circular. Also the logarithmic gradient (e.g.
$-\partial \ln\phi/\partial r$ near $z=0$) rises steeply as we move past the
charge extent in the transverse direction (slightly overshooting and
then) asymptoting to unity. These are universal qualitative features
of such solutions. Of course, close-to or inside the blue (charge) region, the
shape of the contours is strongly modified in accordance with the
shape of the charge perturbation (just illustrative here); and it is
the charge distribution that determines the resulting aspect ratio
$L_\perp/L_\parallel$ there. Oblate holes rely on their proximity to oblate
distributions of trapped electron deficit for their shape, not on a
supposed anisotropy of shielding arising from transverse polarization.

\section{Conclusion}

The empirical aspect ratio scaling of Franz et al
$L_\perp/L_\parallel \simeq \sqrt{1+\omega_p^2/\Omega^2}$ may
represent a useful (though perhaps not universal) empirical fit
to electron hole shapes in nature. But it cannot be explained by the
equations of the ``model'' that they invoke because their anisotropic
Poisson equation is based on a misunderstanding of polarization
drift's role and representation in gyrokinetic theory. It is
shown here on general grounds that we can expect the electron hole
potentials to decay essentially isotropically with local density
approximately Boltzmann-like as $\phi\to 0$, corresponding to
isotropic Debye screening. Charge density arising, for example, from
trapped particle deficit is isotropically screened by untrapped
particles. This is confirmed by simple PIC simulation. The wings of
electron holes therefore are expected to have asymptotically
exponential potential decay both parallel and perpendicular to the
magnetic field.

\subsection*{Acknowledgements}
I am grateful to Greg Hammett for several very helpful discussions on
the foundations of gyrokinetic theory. 

\bibliography{JabRef}

\end{document}